\documentclass[
aps, %
prl,
preprintnumbers,
twocolumn,
superscriptaddress,
nofootinbib,
floatfix,
10pt
]{revtex4-1}

\usepackage{amsfonts,amssymb,stmaryrd,latexsym,amsmath,braket}
\usepackage{graphicx,subfigure}
\usepackage{times}
\usepackage{slashed}
\usepackage{braket}
\usepackage{verbatim}
\usepackage[utf8]{inputenc}

\usepackage{color}
\usepackage[dvipsnames]{xcolor}
\newcommand{\YZ}[1]{{\color{black}#1}}

\usepackage[backref=page]{hyperref}
\hypersetup{pdftitle={SM},
 pdfauthor={YZMA},
 unicode=true, bookmarks=true,bookmarksnumbered=false,bookmarksopen=false, breaklinks=false,pdfborder={0 0 1},backref=false,colorlinks=true, allcolors=BlueViolet}
\usepackage{bm}
\usepackage{soul}

\begin{document}

\title{Structure Factors for Hot Neutron Matter \\ from {\it Ab Initio} Lattice Simulations
with High-Fidelity Chiral Interactions}

\author{Yuan-Zhuo Ma}
\affiliation{%
Key Laboratory of Atomic and Subatomic Structure and Quantum Control (MOE), Institute of Quantum Matter, South China Normal University, Guangzhou 510006, China
}%
\affiliation{%
Facility for Rare Isotope Beams and Department of Physics and Astronomy, Michigan State University, MI 48824, USA
}%

\author{Zidu Lin}
\affiliation{Department of Physics and Astronomy, University of Tennessee Knoxville}

\author{Bing-Nan Lu}
\affiliation{%
Graduate School of China Academy of Engineering Physics, Beijing 100193, China
}%

\author{Serdar Elhatisari}
\affiliation{%
Faculty of Natural Sciences and Engineering, Gaziantep Islam Science and Technology University, Gaziantep 27010, Turkey
}%
\affiliation{%
Helmholtz-Institut für Strahlen- und Kernphysik and Bethe Center for Theoretical Physics, Universität Bonn, D-53115 Bonn, Germany
}%
 
\author{Dean Lee}\thanks{leed@frib.msu.edu}
\affiliation{%
Facility for Rare Isotope Beams and Department of Physics and Astronomy, Michigan State University, MI 48824, USA
}%

\author{Ning Li}
\affiliation{%
School of Physics, Sun Yat-Sen University, Guangzhou 510275, China
}%

\author{Ulf-G. Meißner}
\affiliation{%
Helmholtz-Institut für Strahlen- und Kernphysik and Bethe Center for Theoretical Physics, Universität Bonn, D-53115 Bonn, Germany
}%
\affiliation{%
Institute for Advanced Simulation, Institut für Kernphysik, and Jülich Center for Hadron Physics, Forschungszentrum Jülich, D-52425 Jülich, Germany
}%
\affiliation{%
Tbilisi State University, 0186 Tbilisi, Georgia
}%

\author{Andrew W. Steiner}
\affiliation{Department of Physics and Astronomy, University of Tennessee Knoxville}
\affiliation{Physics Division, Oak Ridge National Laboratory}

\author{Qian Wang}
\affiliation{%
Key Laboratory of Atomic and Subatomic Structure and Quantum Control (MOE), Institute of Quantum Matter, South China Normal University, Guangzhou 510006, China
}%

\begin{abstract}
We present the first \emph{ab initio} lattice calculations of spin and density correlations in hot neutron matter using high-fidelity interactions at next-to-next-to-next-to-leading order (N3LO) in chiral effective field theory.  These correlations have a large impact on neutrino heating and shock revival in core-collapse supernovae and are encapsulated in functions called structure factors. Unfortunately, calculations of structure factors using high-fidelity chiral interactions were well out of reach using existing computational methods.  In this work, we solve the problem using a computational approach called the rank-one operator (RO) method.  The RO method is a general technique with broad applications to simulations of fermionic many-body systems. It solves the problem of exponential scaling of computational effort when using perturbation theory for higher-body operators and higher-order corrections.  Using the RO method, we compute the vector and axial static structure factors for hot neutron matter as a function of temperature and density. The \emph{ab initio} lattice results are in good agreement with virial expansion calculations at low densities but are more reliable at higher densities.  Random phase approximation codes used to estimate neutrino opacity in core-collapse supernovae simulations can now be calibrated with {\it ab initio} lattice calculations.  
\end{abstract}

\maketitle

\paragraph{}{\itshape Introduction.} Core-collapse supernovae (CCSNe) are catastrophic events heralding the death of massive stars.  Under enormous gravitational pressure, the nickel-iron core converts to neutron-rich matter via inverse beta decay.  This results in an infall of stellar matter followed by a violent rebound from the ultradense core.  Meanwhile, copious numbers of neutrinos are produced. 
 This neutrino flux provides energy to the shock wave and increases the likelihood of an explosion.  Since the neutrino-nucleon scattering rates are greatly modified by the spin and density correlations in neutron-rich matter, understanding these correlations is important for modeling CCSNe explosions \cite{Mezzacappa2005, BURROWS2006, Burrows2013, Burrows2021}. 
 Early efforts in studying in-medium neutrino-nucleon scattering have used mean field methods such as the Hartree-Fock (HF) and random phase approximations (RPA) \cite{Sawyer1989, Reddy1999, HOROWITZ2006, Horowitz2017}.
Extended virial expansions provide model-independent predictions in the limit of low densities and high temperatures \cite{HOROWITZ2006,Horowitz2017, Lin2017, Berger2020, Hou2020}.

More recently, {\it ab initio} lattice calculations of neutron matter and its structure factors were performed using pionless effective field theory at leading order, both in the limit of infinite scattering length \cite{Alexandru2020} and at the physical scattering length \cite{Alexandru2021}.  These calculations are suitable for environments where the neutrons have momenta less than $100$~MeV.  We are using natural units where the speed of light, $c$, reduced Planck constant, $\hbar$, and Boltzmann constant, $k_B$, are set to unity.  In order to describe neutron matter at densities and temperatures relevant for CCSNe, a good description of nucleons up to $300$~MeV momenta is needed.  The standard theoretical framework for this regime is provided by chiral effective field theory ($\chi$EFT), where the forces mediated by the exchange of pions are treated explicitly \cite{Epelbaum:2008ga,Machleidt:2011zz}.

Recent advances in chiral effective field theory interactions and advanced quantum many-body methods have pushed forward the frontiers of \emph{ab initio} nuclear calculations.  Calculations are now possible for
light nuclei \cite{LEE2009, Epelbaum2012, Pastore:2017uwc, BARRETT2013131, Hupin:2018biv, Carlson2015}, medium-mass \cite{Hagen_2014, LU2019,Soma:2020dyc,Wirth:2021pij,Stroberg2021}, and heavy nuclei \cite{Hu:2021trw}, nuclear matter \cite{Lynn2016, Buraczynski2016, Buraczynski2019,Drischler2021} as well as finite temperature systems \cite{Mukherjee2007,Soma2009,Carbone2019,Lu2020,Carbone2020,Keller:2022crb}.  In this work, we compute the spin and density correlations in neutron matter at various temperatures and densities using lattice chiral effective field theory at next-to-next-to-next-to leading order (N3LO). Such N3LO lattice calculations were previously not possible due to the many perturbation theory corrections required and the lack of a practical method for computing the corrections efficiently.

In this work, we introduce a new computational approach called the rank-one operator (RO) method.  In many-body theory, rank-one operators are one-body operators where one creation operator multiplies one annihilation operator.  As we will show, rank-one operators are special since they can be inserted into auxiliary-field Monte Carlo calculations without the need to compute derivatives with respect to parameters.  The RO method uses this property of rank-one operators to solve the problem of exponential scaling of computational effort when using perturbation theory for higher-body operators and higher-order corrections.  It can be used with Monte Carlo calculations of fermionic systems in nuclear physics, condensed matter, ultracold atomic gases, and quantum chemistry.

\paragraph{}{\itshape Methods.} Nuclear lattice effective theory (NLEFT) is an \emph{ab initio} method that combines effective field theory with lattice Monte Carlo (MC) simulations \cite{Lee2007,LEE2009, Lahde:2019npb, elhatisari_2015, Elhatisari2016, Li2016, Elhatisari2017, LU2019, Lu2020, Lu2022}.
The use of unrestricted Monte Carlo simulations allows for investigations of strong many-body correlations such as clustering \cite{Elhatisari2017, Nicholas2021, Shen:2022bak}.  Moreover, the pinhole trace algorithm \cite{Lu2020} allows for \emph{ab initio} calculations of nuclear thermodynamics.  

For fixed neutron number $N$ and temperature $T$, the expectation value of an observable ${O}$ in the canonical ensemble (CE) is given by
\begin{equation}\label{eq:1}
    \langle{O}\rangle_{N}=\frac{Z_{\mathcal{O}}(\beta, N)}{Z( \beta, N)}=\frac{\operatorname{Tr}_{N}\left(e^{-\beta H} {O}\right)}{\operatorname{Tr}_{N}\left(e^{-\beta H}\right)},
\end{equation}
where $\beta = T^{-1}$ is the inverse of temperature, $H$ is the Hamiltonian, and $\text{Tr}_N$ is the trace over all the $N$-neutron states.
The canonical partition function $Z(\beta, N)$, can be written explicitly in the single-particle basis $c_{i}=\left(\boldsymbol{n}_{i}, \sigma_{i}, \tau_{i}\right)$ as
\begin{equation}\label{eq:2}
    Z(\beta, N)=\sum_{c_{1}, \ldots, c_{N}}\left\langle c_{1}, \ldots, c_{N}|\exp (-\beta H)| c_{1}, \ldots, c_{N}\right\rangle,
\end{equation}
with $\boldsymbol{n}_{i}$ an integer triplet specifying the lattice coordinates, $\sigma_{i}$ is the spin, and \YZ{$\tau_{i}=-{1}/{2}$ is the isospin for neutrons}.
To update and sum over initial/final states we implement the pinhole trace algorithm described in Ref.~\cite{Lu2020}. 

We break up the exponential $\exp(-\beta H)$ as a product of transfer matrices, which are just short-time exponentials for each lattice time step.   In the auxiliary field formalism, the transfer matrices depend on the auxiliary fields and pion fields \cite{LEE2009,Lahde:2019npb}.  The transfer matrix $M(n_t)$ corresponds to time step $n_t$.  If we use $L_t$ total time steps, then we get a product of the form $M(L_t-1) \cdots M(0)$. We use the shuttle algorithm described in Ref.~\cite{LU2019} to update the auxiliary fields and pion fields.

It is also convenient to consider the partition function for the grand canonical ensemble (GCE), 
\begin{equation}\label{eq:Z_G}
        \begin{aligned} 
            \mathcal{Z}(\beta,\mu_G) = \sum_{N} e^{\beta \mu_G N } Z(\beta,N),
        \end{aligned}
\end{equation} 
where $\mu_G$ is chemical potential.  
The expectation of an operator in the GCE can be evaluated as 
\begin{equation}\label{eq:GCE}
        \begin{aligned}
            \langle {O}\rangle_{G} 
            = \frac{ \sum_N \langle {O}\rangle_{N} e^{\beta\mu_G N } Z(\beta,N) }{\sum_N e^{\beta \mu_G N } Z(\beta,N)}
            = \sum_N \langle {O}\rangle_{N} w_{N},
        \end{aligned}
\end{equation}
where $w_{N}$ is the normalized neutron number probability.
This distribution function $w_{N}$ can be obtained by calculating CE partition function $Z(\beta,N) = e^{- \beta F(\beta, N)}$, where the free energy $F(\beta, N)$ is the integration of the CE chemical potential $\mu(\beta, n)$ from an $N_0$-particle system, $F(\beta, N) = F(\beta, N_0) + \int_{N_0}^N \mu (\beta,n) dn $.
We use the Widom insertion method \cite{Widom1963, Binder1997, Lu2020} to calculate the CE chemical potential $\mu(\beta, n)$.
Further details are presented in the Supplemental Material.

A primary challenge for NLEFT calculations is the Monte Carlo sign problem, caused by cancellations between positive and negative amplitudes.  In order to mitigate this problem, we start from a simple interaction with no significant sign oscillations and use the perturbation theory to implement the difference between the simple interaction and the high-fidelity interaction. Perturbative calculations up to the second order correction in the energy have been implemented in lattice quantum Monte Carlo (QMC) simulations~\cite{Lu2022}.
In this work we perform \YZ{the first order perturbation to bridge the gap between simple and high-fidelity chiral interactions}, however we greatly accelerate the convergence of perturbation theory using high-fidelity interactions generated using the method of wave function matching as described in Ref.~\cite{Elhatisari2022}.

In the auxiliary field formalism, we work with Slater determinant wave functions and transfer matrices $M(n_t)$ that consist of exponentials of one-body operators that are normal ordered so that annihilation operators are on the right and creation operators are on the left.  As a result, the many-body amplitude equals the matrix determinant of the single-nucleon amplitudes.  At zeroth order in perturbation theory, we simply replace each $M(n_t)$ by the unperturbed transfer matrix $M^{(0)}(n_t)$.  In order to calculate perturbation theory corrections, we introduce additional terms into the transfer matrices,
\begin{equation}
    M(n_t) = M^{(0)}(n_t) + \sum_{\YZ{\theta}} t_{\YZ{\theta}}(n_t)O_{\YZ{\theta}} \cdots, \label{eq:perturb_insert}
\end{equation}
where each $O_{\YZ{\theta}}$ is a normal-ordered one-body operator.  We can insert the operator $O_{\YZ{\theta}}$ wherever desired by taking the derivative with respect to the corresponding parameter $t_{\YZ{\theta}}(n_t)$ and setting all such parameters to zero thereafter.  In this manner, we can build higher-body operators from products of the one-body operators $O_{\YZ{\theta}}$ and compute corrections to any order in perturbation theory.

Unfortunately, the severe computational challenge one faces is that taking $k$ such derivatives requires $O(2^k)$ terms.  The exponential scaling is readily seen when one calculates the derivatives using finite differences.  Each finite difference requires a forward and backward step, and this produces $2^k$ terms for $k$ derivatives.  The Jacobi formulas for derivatives of matrix determinants \cite{Lahde:2019npb} allows us to calculate the derivatives exactly without finite differences, however the scaling of computational effort is still $O(2^k)$.

The RO method avoids this exponential scaling by using one-body operators $O_{\YZ{\theta}}$ that have the form $F^\dagger_{\alpha'}F_{\alpha}$, where $F_{\alpha}$ is the annihilation operator for nucleon orbital $\alpha$ and $F^\dagger_{\alpha'}$ is the creation operator for nucleon orbital $\alpha'$.  Since $F_{\alpha}$ can only annihilate one nucleon and $F^\dagger_{\alpha'}$ can only create one nucleon, it is a rank-one operator.  We conclude that the amplitude has no terms that contain more than one power of the coefficient $t_{\YZ{\theta}}(n_t)$.  Instead of inserting $O_{\YZ{\theta}}$ by taking the derivative with respect to $t_{\YZ{\theta}}(n_t)$, we can simply take $t_{\YZ{\theta}}(n_t)$ to be very large and divide the amplitude by $t_{\YZ{\theta}}(n_t)$.  Since this requires the calculation of one amplitude rather than two amplitudes, the problem of exponential scaling is solved.

Static structure factors are Fourier transforms of the fluctuations of the spin and density correlation functions.  
Let $\hat{\rho}$ and $\hat{\rho}_z$ be the particle density and spin density operators, respectively.  Let $\rho^0$ be the average particle density and $\rho_z^0$ be the average spin density.  Here we consider unpolarized neutron matter where $\rho_z^0$ equals zero.  On the lattice, the vector and axial static structure factors can be written as
\begin{equation} \label{eq:Sva}
        \begin{aligned}
        & S_{\rm{v}}(\bm{q})=\frac{1}{L^3}\sum_{\bm{n} \bm{n}'} \mathrm{e}^{-i \bm{q} \cdot \bm{n}}\left[\langle \hat{\rho}(\bm{n} + \bm{n}') \hat{\rho}(\bm{n}')\rangle-(\rho^0)^2\right], \\
        & S_{\rm{a}}(\bm{q})=\frac{1}{L^3}\sum_{\bm{n} \bm{n}'} \mathrm{e}^{-i \bm{q} \cdot \bm{n}}\left[\langle \hat{\rho}_z(\bm{n} + \bm{n}') \hat{\rho}_z(\bm{n}') \rangle - (\rho_z^0)^2 \right],
        \end{aligned}
\end{equation}
where $\boldsymbol{n}, \boldsymbol{n}'$ represents coordinate on a $L^3$ cubic lattice.  The expectation values of these two-body density correlation operators in Eq.~\eqref{eq:Sva} are calculated using the RO formalism.  Further details can be found in Supplemental Materials \cite{SM}. 

\paragraph{\itshape Results.} We perform simulations on $L^3=6^3, 7^3, 8^3$ cubic lattices with spatial lattice spacing $a = 1/(150 \text{ MeV}) \approx 1.32$~fm and temporal lattice spacing $a_t = 1/(1000 \text{ MeV})$.
Following the strategy in Ref. \cite{Lu2020}, we use twist-averaged boundary conditions to eliminate finite volume effects and accelerate the convergence to the thermodynamic limit.
For twist angle $\theta_i$ along each spatial direction $i$, the possible lattice momenta are $2\pi{n}_i/L + \theta_i/L$, with some integer $n_i$ .
The averaging over all possible twist angles $\theta_i$ is done by Monte Carlo sampling.

\begin{figure}[h] 
\centering 
\includegraphics[width=0.48\textwidth]{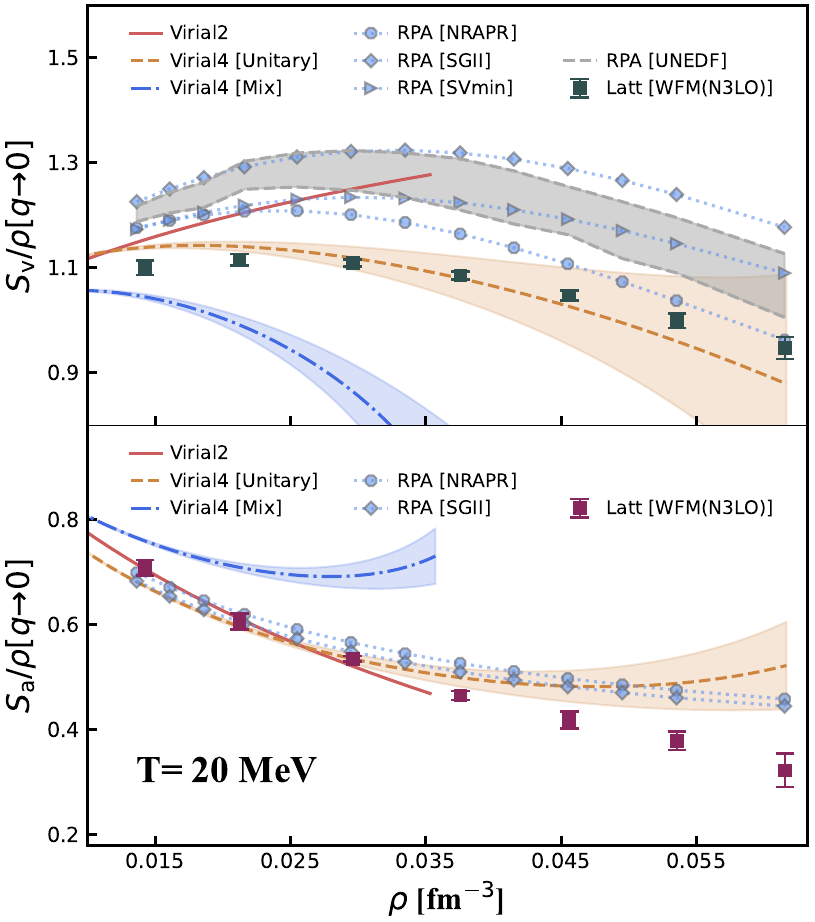} 
\caption{Calculated static structure factors of $S_{\rm{v}}$ and $S_{\rm{a}}$ at the long-wavelength limit ($q\to0$) with $T=20$ MeV.
Virial2 denotes 2nd-order virial calculations using physical neutron data.  Virial4 [Unitary] corresponds to 4th virial calculations for the unitary limit. Virial4 [Mix] is a hybrid of the two, with the 2nd-order term for physical neutrons and the 3rd- and 4th-order terms for the unitary limit.  The RPA calculations are carried out with four different interactions (NRAPR, SGII, SVmin and UNEDF). 
WFM(N3LO) represents the NLEFT calculations with the wave function matching N3LO interaction.
} 
\label{fig:1} 
\end{figure}

\paragraph{} In Fig.~\ref{fig:1} we present NLEFT results in the GCE for the static structure factors $S_{\rm{v}}$ and $S_{\rm{a}}$ in the long wavelength limit, $q \rightarrow 0$.  The results are calculated at a temperature of $20$~MeV and plotted as a function of density. 
We show lattice results corresponding to the high-fidelity N3LO chiral interaction generated using wave function matching (WFM) \cite{Elhatisari2022}.  
For comparison, we show results obtained with RPA calculations using NRAPR \cite{Steiner:2004fi}, SGII \cite{vanGiai:1981zz}, SVmin \cite{Reinhard:2010wz} and UNEDF \cite{McDonnell:2015sja} Skyrme interactions.  We also show several virial expansion calculations, which we now discuss.

The virial expansion is an expansion in powers of the fugacity $z=\exp(\mu/T)$.  We also make use of an approximate idealization of pure neutron matter called the unitary limit, where the interaction range is zero and the the scattering length is infinite.  The results labelled as Virial2 in Fig.~\ref{fig:1} corresponds to the virial expansion at 2nd order, using physically observed data for the interactions between neutrons \cite{Horowitz2017}.  The Virial4 [Unitary] results show virial expansions for the unitary Fermi gas results at 4th order \cite{Lin2017}.  Higher-order virial coefficients corresponding to physical neutrons are not currently available.  The Virial4 [Mix] results corresponds to a hybrid virial calculation where the 2nd-order terms correspond to physical neutrons but the 3rd- and the 4th-order terms are associated with the unitary limit. The error bands on the Virial4 results are associated with uncertainties in the 4th-order virial coefficient in the unitary limit.  The significant difference between Virial4 [Unitary] and Virial4 [Mix] shows that even a minor change to the interaction has a significant effect on the vector and axial static structure factors.

In the very low-density region, the lattice results are in agreement with both Virial2 and Virial4 [Mix].  For larger densities, the wide difference between Virial2 and Virial4 [Mix] shows that the order-by-order convergence of the virial expansion is slow. Further details are discussed in the Supplemental Material.  The Virial4 [Unitary] results intersect with the lattice results near density $0.030$~fm$^{-3}$. 
\YZ{However, the deviations with the lattice results can be as large as $\sim 25 \%$ at $n \approx 0.053 ~\mathrm{fm}^{-3}$ and $\sim 5\%$ at $n\approx 0.015 ~\mathrm{fm}^{-3}$ for $S_{\rm{a}}$. }

The RPA calculations provide self-consistent yet model-dependent calculations of structure factors not only for pure neutron matter but also for beta-equilibrium matter in a wide range of densities and temperatures \cite{Lin:2022lug}.  The UNEDF interaction has quantified uncertainties of the Skyrme interactions, and we generate an error band of RPA $S_{\rm{v}}$ corresponding to these uncertainties. 
In the axial current channel, we present RPA calculations for NRAPR and SGII, interactions for which the problem of Skyrme interaction spin instabilities do not appear for densities lower than the saturation density of nuclear matter. In both the vector and axial current channels, we note that the structure factors for some of the RPA calculations are in reasonable agreement with the lattice calculations. 
In the Supplemental Material, we use the lattice results to calibrate the Skyrme interactions used in the RPA calculations and make predictions for the neutrino inverse mean free path. Several corresponding results for the chemical potential, fugacity, density and pressure are listed in Table~\ref{tab:1}. The uncertainties shown in the graphs and the table are stochastic errors only.  We discuss systematic errors at the end of this section.
\paragraph{}


\begin{table}
\caption{\label{tab:1} Calculated grand canonical ensemble fugacity $z$, density $\rho$ [fm$^{-3}$], pressure $p$ [MeV/fm$^3$] with different chemical potential $\mu$ [MeV]  at $T=20$ MeV.}
\centering{}%
\setlength{\tabcolsep}{0.12cm}
\begin{tabular}{ccccccc}
\hline
\hline
$\mu$  & $-$23.78 &  $-$16.58 &$-$10.787 &$-$6.450 & $-$2.828 \tabularnewline
$z$    & 0.3045   & 0.4365    & 0.5831   & 0.7243  & 0.8682   \tabularnewline
 $\rho$  $\times 100$  
       & 1.4274(4) & 2.122(1) & 2.961(1) & 3.758(1) & 4.560(1) \tabularnewline
$p$  & 0.2458(1) & 0.3618(1)  &0.4901(1) & 0.6287(1) & 0.7737(2)\tabularnewline
\hline
\hline
\end{tabular}
\end{table}

\begin{figure}[h] 
\centering 
\includegraphics[width=0.48\textwidth]{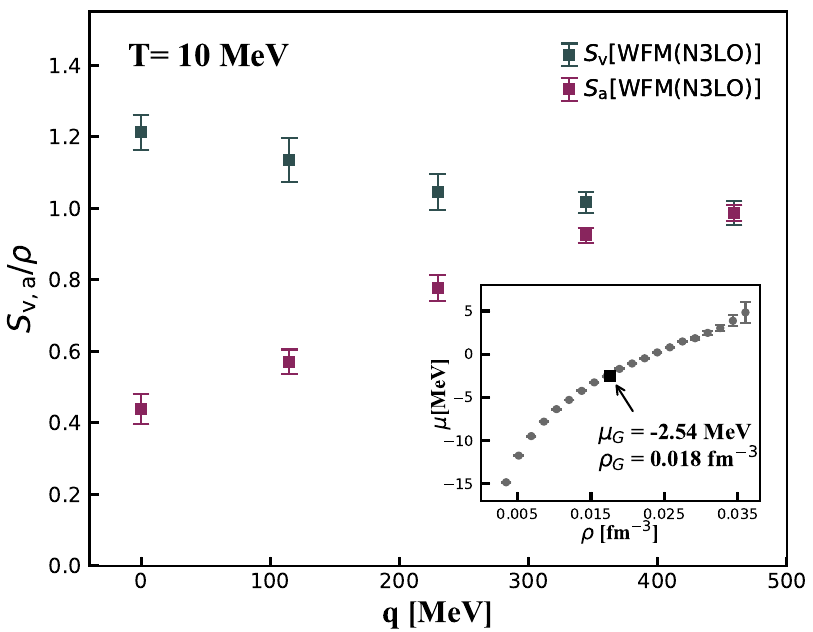} 
\caption{Calculated momentum dependent neutron matter structure factors $S_{\rm{v}}$ and $S_{\rm{a}}$ at $T=10$ MeV.
WFM(N3LO) represents the NLEFT calculations with the wave function matching N3LO interaction. 
The insert figure shows calculated chemical potentials of CE systems which are used for the construction of GCE at the chemical potential $\mu_G = -2.54$~MeV and the density $\rho_G = 0.01758 (4)$~fm$^{-3}$.
} 
\label{fig:2} 
\end{figure}
In Fig.~\ref{fig:2}, we present the momentum-dependent structure factor calculations at a temperature of $10$ MeV in the GCE at density $\rho_G = 0.018$ fm$^{-3}$.
The lattice cutoff momentum is $\pi/a = 470$ MeV, and so the lattice results are most reliable for momenta smaller than this momentum scale.  Nevertheless, the lattice results show the expected high-momentum behavior.  Both $S_{\rm{v}}$ and $S_{\rm{a}}$ should equal the system density at large momenta, $S_{\rm{v}}(q \rightarrow \infty)=S_{\rm{a}}(q \rightarrow \infty)=\rho$.

At long wavelengths, $S_{\rm{v}}$ and $S_{\rm{a}}$ have opposite trends.
We also present the calculated chemical potentials of 20 CE systems in the inset of Fig.~\ref{fig:2}, which are used to construct this GCE system.  
The many-body corrections on the neutral current neutrino-nucleon interactions in CCSNe are usually estimated in the long wavelength limit, which is justified since the typical momentum transfer by scattered neutrinos is small compared to the thermal nucleon momentum $\sqrt{6MT}$, where $M$ is the nucleon mass. 
Note by applying exact dynamic structure factors in the calculation, the neutral current neutrino-nucleon scattering rates have small but noticeable deviations from the ones estimated in long wavelength limit \cite{Bedaque:2018wns}. Our \emph{ab initio} calculations at finite momentum transfer provide benchmarks for the calculation of dynamic structure factors of pure neutron matter at finite temperatures. 

In addition to the statistical errors reported for the lattice results, we estimate an overall systematic uncertainty at the 5\% level. The largest sources of systematic errors are due to finite system size errors, uncertainties in the nuclear interaction, and an approximation made in neglecting the numerically small higher-order corrections to the chemical potential, as discussed in the Supplemental Material.  The finite system size error was obtained by analyzing the density and pressure with different box sizes.  The errors due to uncertainties in the nuclear interaction and the neglected higher-order corrections to the chemical potential were performed by comparing results from calculations at different chiral orders in the canonical ensemble.

\paragraph{Summary.}  
We have performed the first \emph{ab initio} calculation of structure factors for hot neutron matter using high-fidelity chiral interactions at N3LO.  The lattice results of vector and axial structure factors are in good agreement with virial expansions at low densities. The lattice predictions as a function of density, temperature, and momentum transfer provide valuable benchmarks for calibrating RPA and other models commonly used in supernovae simulations.  This is detailed in Supplemental Materials and further studies are planned in the future.

We have introduced a new computational approach called the rank-one operator method to perform the calculations presented in this work. The rank-one operator method should have immediate applications to Monte Carlo simulations for nearly any quantum many-body systems composed of fermions.  With new technologies such as wave function matching available to accelerate the convergence of perturbation theory, one has the possibility of avoiding Monte Carlo sign problems for a large class of fermionic many-body systems.  The new computational paradigm requires computing amplitudes with multiple insertions of higher-body operators, and the rank-one operator method is ideally suited for this purpose.

\begin{acknowledgements}
\paragraph{Acknowledgement}
We are grateful for useful discussions with Chuck Horowitz, Gustav Jansen and members of the Nuclear Lattice Effective Field Theory Collaboration.
This work has been supported by the Guangdong Major Project of Basic and Applied Basic Research No. 2020B0301030008, the National Natural Science Foundation of China (Grants No.~12105106, No.~12035007 and No.~12275259), China Postdoctoral Science Foundation under Grant No. BX20200136 and 2020M682747; ZL and AWS were supported by NSF PHY 21-16686. AWS was also supported by the Department of Energy Office of Nuclear Physics. The work of BNL was supported by NSAF No.U2330401. DL was supported in part by the U.S. Department of Energy (DE-SC0021152, DE-SC0013365, DE-SC0023658, SciDAC-5 NUCLEI Collaboration).
The work of UGM was supported by the European Research Council (ERC) under the European Union's Horizon 2020 research and innovation programme (grant agreement No. 101018170), by Deutsche Forschungsgemeinschaft (DFG, German Research Foundation) (project-ID 196253076 - TRR 110), the Chinese Academy of Sciences (CAS) President's International Fellowship Initiative (PIFI) (Grant No. 2018DM0034), Volkswagen Stiftung (Grant No. 93562) and by the MKW NRW (funding code NW21-024-A). The work of QW was supported by Guangdong Provincial funding with Grant No.~2019QN01X172.
Computational resources provided by the Oak Ridge Leadership Computing Facility through the INCITE award
``Ab-initio nuclear structure and nuclear reactions'', the Southern Nuclear Science Computing Center in the South China Normal University, the Gauss Centre for Supercomputing e.V. (www.gauss-centre.eu) for computing time on the GCS Supercomputer JUWELS at the J{\"u}lich Supercomputing Centre (JSC), and the Institute for Cyber-Enabled Research at Michigan State University.
\end{acknowledgements}

\nocite{reinert_semilocal_2018}
\nocite{Li:2018ymw}
\nocite{PWA}
\nocite{Lee2021}
\nocite{Lahde:2015ona}
\nocite{PhysRevLett.105.070402}

\bibliography{References}

\bibliographystyle{apsrev} 

\end{document}



\section*{SUPPLEMENTAL MATERIAL}
\addtocounter{section}{10}



\subsection{Wave function matching and $\chi$EFT Hamiltonian} 
In our lattice calculations, we use wave function matching \cite{Elhatisari2022} and perturbation theory to mitigate the Monte Carlo ``sign problem''.
Starting from the high-fidelity $\chi$EFT Hamiltonian $H$, wave function matching performs a unitary transformation to create a new high-fidelity Hamiltonian $H'$ such that wave functions at short distances match that of a simple Hamiltonian $H^S$.
This unitary transformation can provide a rapidly converging expansion in powers of the difference $H' - H^S$.
It has been shown that a Hamiltonian with Wigner's SU(4) symmetry will be positive definite and has no sign problem \cite{Lee2007}.
Thus, the simple Hamiltonian $H^{S}$ can be constructed with approximate SU(4) symmetry, and the gap $H' - H^S$ can be filled by lattice perturbation theory. 

We choose a leading-order $\chi$EFT interaction for the simple Hamiltonian \cite{Elhatisari2022, Lu2022},
\begin{equation}\label{eq:HS}
        H^S=K+\frac{1}{2} c_{\rm{SU4}} \sum_{\bm{n}}: \tilde{\rho}^2(\bm{n}) :+V_{\mathrm{OPE}}^{\Lambda_\pi},
\end{equation}
where $K$ is the kinetic term with nucleon mass $m=938.92$ MeV and the $::$ symbols mean normal ordering\DL{, where annihilation operators are on the right and creation operators are on the left.}  $\tilde{\rho}$ is the density operator for nucleons with local and non-local smearing,
\begin{equation}
        \tilde{\rho}(\bm{n})=\sum_{i, j=0,1} \tilde{a}_{i, j}^{\dagger}(\bm{n}) \tilde{a}_{i, j}(\bm{n})+s_{\mathrm{L}} \sum_{\left|\bm{n}-\bm{n}^{\prime}\right|=1} \sum_{i, j=0,1} \tilde{a}_{i, j}^{\dagger}\left(\bm{n}^{\prime}\right) \tilde{a}_{i, j}\left(\bm{n}^{\prime}\right).
\end{equation}
The non-locally smeared annihilation and creation operators, $\tilde{a}$ and $\tilde{a}^{\dagger}$, with spin $i=0,1$ (up, down) and isospin $j=0,1$ (proton, neutron) indices are defined as,
\begin{equation}
        \tilde{a}_{ij}(\boldsymbol{n})=a_{ij}(\boldsymbol{n})+s_{\rm{N L}} \sum_{\left|\boldsymbol{n}^{\prime}-\boldsymbol{n}\right|=1} a_{ij}\left(\boldsymbol{n}^{\prime}\right).
\end{equation}
We use local smearing parameter $s_{\rm{L}} = 0.07$ and non-local smearing parameter $s_{\rm{NL}} = 0.5$. 

In addition to the short-range SU(4) symmetric interaction, there is also the one-pion-exchange (OPE) potential appearing at the leading order,
\begin{equation}\label{eq:OPE}
        V_{\mathrm{OPE}}^{\Lambda_\pi}
        =-\frac{g_A^2}{8 F_\pi^2} \sum_{\bm{n}', \bm{n}, S^{\prime}, S, I}: \rho_{S^{\prime}, I}\left(\bm{n}^{\prime}\right) f_{S^{\prime}, S}\left(\bm{n}^{\prime}-\bm{n}\right) \rho_{S, I}(\bm{n}): ,
\end{equation}
\begin{equation}\label{eq:OPEcounter}
    V_{\mathrm{C}_\pi}^{\Lambda_\pi} = 
        -C_\pi \frac{g_A^2}{8 F_\pi^2} \sum_{\mathbf{n}^{\prime}, \mathbf{n}, S, I}: \rho_{S, I}\left(\bm{n}^{\prime}\right) f^\pi\left(\bm{n}^{\prime}-\bm{n}\right) \rho_{S, I}(\bm{n}): ,
\end{equation}
where $C_\pi$ is defined as,
\begin{equation}
        C_\pi=-\frac{\Lambda_\pi\left(\Lambda_\pi^2-2 M_\pi^2\right)+2 \sqrt{\pi} M_\pi^3 \exp \left(M_\pi^2 / \Lambda_\pi^2\right) \operatorname{erfc}\left(M_\pi / \Lambda_\pi\right)}{3 \Lambda_\pi^3}.
\end{equation}
We regularize the OPE potential by a Gaussian form factor in momentum space~\cite{reinert_semilocal_2018}. Here $f^{\pi}$ is a local regulator in momentum space,
\begin{equation}
        f^\pi\left(\bm{n}^{\prime}-\bm{n}\right)=\frac{1}{L^3} \sum_{\bm{q}} e^{-i \bm{q} \cdot\left(\bm{n}^{\prime}-\bm{n}\right)-\left(\bm{q}^2+M_\pi^2\right) / \Lambda_\pi^2},
\end{equation}
and $f_{S', S}$ is the locally-regulated pion correlation function,
\begin{equation}
        f_{S^{\prime}, S}\left(\bm{n}^{\prime}-\bm{n}\right)=\frac{1}{L^3} \sum_{\bm{q}} \frac{q_{S^{\prime}} q_S e^{-i \bm{q} \cdot\left(\bm{n}^{\prime}-\bm{n}\right)-\left(\bm{q}^2+M_\pi^2\right) / \Lambda_\pi^2}}{\bm{q}^2+M_\pi^2},
\end{equation}
where $L$ is the length of our cubic box and momentum components $q_{S}$ on the lattice are integers multiplied by $2\pi/L$. Finally, $\rho_{S,I}$ is the spin- and isospin-dependent density operator for nucleons, 
\begin{align}
\rho_{S,I}(\mathbf{n}) = & \sum_{i,j,i^{\prime},j^{\prime}=0,1} 
\tilde{a}^{\dagger}_{i,j}(\mathbf{n}) \, [\sigma_{S}]_{ii^{\prime}} \, [\sigma_{I}]_{jj^{\prime}} \, \tilde{a}^{\,}_{i^{\prime},j^{\prime}}(\mathbf{n})
\nonumber \\
& +
s_{\rm L}
 \sum_{|\mathbf{n}-\mathbf{n}^{\prime}|= 1} 
 \,
  \sum_{i,j,i^{\prime},j^{\prime}=0,1} 
\tilde{a}^{\dagger}_{i,j}(\mathbf{n}^{\prime}) \, [\sigma_{S}]_{ii^{\prime}} \, [\sigma_{I}]_{jj^{\prime}} \, \tilde{a}^{\,}_{i^{\prime},j^{\prime}}(\mathbf{n}^{\prime})\,,
\end{align}
where $\sigma_{S}$ are Pauli matrices in spin space and $\tau_{I}$ are Pauli matrices in isospin space.

In the equations above, $g_A = 1.287$ is the axial-vector coupling constant (corrected for the Goldberger-Treiman discrepancy), $F_{\pi} = 92.2$~MeV is the pion decay constants, and $M_{\pi} = 134.98$~MeV is the (neutral) pion mass.
The interaction given in Eq.~\eqref{eq:OPEcounter} is a counterterm introduced to remove the short-range singularity from the one-pion exchange potential.
In the simple Hamiltonian, we set $\Lambda_{\pi} = 180$~MeV and $C_{\pi} = 0$, and we treat the difference $V_{\mathrm{OPE}}^{\Lambda_\pi = 300~{\rm MeV}} - V_{\mathrm{OPE}}^{\Lambda_\pi=180~{\rm MeV}}$ and the OPEP counterterm $V_{\mathrm{C}_\pi}^{\Lambda_\pi}$ in perturbation theory where $\Lambda_{\pi} = b_{\pi}^{-1/2}$. 
More details can be found in Refs.~\cite{Li:2018ymw,Elhatisari2022,Lu2022}.

The high-fidelity $\chi$EFT Hamiltonian at N3LO has the form,
\begin{equation}\label{eq:wfm}    H=K+V_{\mathrm{OPE}}^{\Lambda_\pi}+V_{\mathrm{C}_\pi}^{\Lambda_\pi}+V_{\mathrm{Cou}}+V_{3\rm{N}}^{\mathrm{Q}^3}+V_{2\rm{N}}^{\mathrm{Q}^4}+W_{2\rm{N}}^{\mathrm{Q}^4}+V_{2\rm{N}, \rm{WFM}}^{\mathrm{Q}^4}+W_{2\rm{N}, \rm{WFM}}^{\mathrm{Q}^4}
\end{equation}
where $V_{\mathrm{Cou}}$ is the Coulomb potential, 
$V_{3\rm{N}}^{\mathrm{Q}^3}$ is the three-body potential, 
$V_{2\rm{N}}^{\mathrm{Q}^4}$ corresponds to the two-body short-range interactions at N3LO, 
$W_{2\rm{N}}^{\mathrm{Q}^4}$ gives the two-body Galilean invariance restoration (GIR) interactions at N${}^3$LO,
$V_{2\rm{N}, \rm{WFM}}^{\mathrm{Q}^4}$ is the wave function matching interaction, 
and $W_{2\rm{N}, \rm{WFM}}^{\mathrm{Q}^4}$ is the GIR correction to the wave function matching interaction.
It should be mentioned that the contribution from $W_{2\rm{N}, \rm{WFM}}^{\mathrm{Q}^4}$ is negligible while the computational cost is large, and we do not include it in our neutron matter simulations.

\YZ{We perform our calculations using lattice spacing $a = 1.32$ fm, and we determine the low-energy constants (LECs) of the 2N short-range interaction up to N3LO of $\chi$EFT by reproducing the neutron-proton scattering phase shifts and mixing angles of the Nijmegen partial wave analysis (PWA) \cite{PWA}. The lattice spacing of $a = 1.32$ fm corresponds to the momentum space cutoff of 470 MeV, which corresponds to the resolution scale at which the hidden spin-isospin symmetry of the NN interactions is best fulfilled \cite{Lee2021}}.
More details are discussed in the Supplemental Material of Ref.~\cite{Elhatisari2022}.

\DL{
\subsection{Jacobi formulas}

In the auxiliary field formalism, the transfer matrices $M(n_t)$ consist of normal-ordered exponentials of one-body operators and the resulting wave functions are Slater determinants.  The many-body amplitude therefore equals the matrix determinant of the single-nucleon amplitudes.  At zeroth order in perturbation theory, we replace each $M(n_t)$ by the unperturbed transfer matrix $M^{(0)}(n_t)$.  When calculating perturbation theory corrections, we introduce additional terms into the transfer matrices,
\begin{equation}
    M(n_t) = M^{(0)}(n_t) + \sum_\theta t_\theta(n_t)O_\theta \cdots, \label{eq:perturb_insert}
\end{equation}
where each $O_\theta$ is a normal-ordered one-body operator.  We can now insert the operator $O_\theta$ anywhere desired by taking the derivative with respect to the corresponding parameter $t_\theta(n_t)$.

Let ${\cal M}$ be the matrix of single-nucleon amplitudes without any operator insertions.  Let ${\cal M}[O_\theta]$ be the new matrix of single-nucleon amplitudes we obtain by inserting $O_\theta$ at some time step $n_t$.  In the following, it is convenient to work with the normal-ordered exponential $:\exp(tO_\theta):$.  The Jacobi identity is a general formula for the derivative of the determinant of a matrix, 
\begin{equation}
	\frac{d}{d t}[\operatorname{det} A(t)]=\operatorname{det} A(t) \cdot \operatorname{tr} [ A^{-1}(t) \cdot \frac{d}{d t} A(t) ].
\end{equation}
Using the Jacobi formula, we have
\begin{equation}\label{eq:jaco_1}
\begin{aligned}
	\det \mathcal{M}[O_\theta]
	&= \frac{d \det \mathcal{M}[:\exp (t {O_\theta}):]}{d t } \Big{|}_{t=0} 
 = \det \mathcal{M} \, \text{tr} \big{\{}  \mathcal{M}^{-1}   \mathcal{M}[O_\theta] \big{\}}.
\end{aligned}
\end{equation}
For the normal-ordered two-body operator, $:O_\alpha O_\beta:$, we use the normal-ordered exponential $:\exp\{t_\alpha O_\alpha + t_\beta {O}_\beta\}:$ and calculate derivatives with respect to $\alpha$ and $\beta$.  The second-order Jacobi formula can be used,
\begin{equation} \label{eq:jaco_2}
	\partial_{t_\alpha}\partial_{t_\beta}[\operatorname{det} A]
        =\operatorname{det} A
	{\left[\operatorname{tr}\left(A^{-1} \partial_{t_\alpha} A\right) \operatorname{tr}\left(A^{-1} \partial_{t_\beta} A\right)-\operatorname{tr}\left(A^{-1} \partial_{t_\alpha} A A^{-1} \partial_{t_\beta} A\right)+\operatorname{tr}\left(A^{-1} \partial_{t_\alpha}\partial_{t_\beta} A\right)\right]}.
\end{equation}
An analogous procedure can be used for second-order perturbation theory where we insert $O_\alpha$ and $O_\beta$ at two different locations.  In that case, $:\exp\{\alpha O_\alpha\}:$ and $:\exp\{\beta {O}_\beta\}:$ are inserted at different time steps, but the same second-order Jacobi formula applies.  In the auxiliary-field formalism, higher-body operators and higher-order terms in perturbation theory are treated in a similar manner.

Unfortunately, the higher-order Jacobi formulas grow exponentially in complexity.  Let $k$ be the total number of one-body operator insertions.  At order $k$, the number of amplitudes that need to be computed scales as $O(2^k)$.  If we use finite differences instead to compute the derivatives numerically, we also get $O(2^k)$ scaling for the number of amplitudes required.  In this work, we need to insert a two-body operator for the structure factor observable as well as two- and three-body $\chi$EFT operators at first-order in perturbation theory.  Such calculations require up to $k = 5$ one-body operator insertions and cannot be performed with high accuracy using current computational resources.  

\subsection{Rank-one operator method}

The RO operator avoids this exponential scaling by using one-body operators $O_\theta$ that have the form $F^\dagger_{\alpha'}F_{\alpha}$, where $F_{\alpha}$ is the annihilation operator for nucleon orbital $\alpha$ and $F^\dagger_{\alpha'}$ is the creation operator for nucleon orbital $\alpha'$.  Since $F_{\alpha}$ can only annihilate one nucleon and $F^\dagger_{\alpha'}$ can only create one nucleon, it is an operator with rank one.  We conclude that the insertion of normal-order exponential $:\exp(t F^\dagger_{\alpha'}F_{\alpha}):$ yields
\begin{equation}
    {\cal M}[:\exp(t F^\dagger_{\alpha'}F_{\alpha}):] = C + t{\cal M}[F^\dagger_{\alpha'}F_{\alpha}].
\end{equation}
The absence of higher-order powers of $t$ allows us compute the ${\cal M}[F^\dagger_{\alpha'}F_{\alpha}]$ very easily by taking the limit of large $t$ and dividing by $t$,
\begin{equation}
    {\cal M}[F^\dagger_{\alpha'}F_{\alpha}] = \lim_{t\rightarrow \infty} \frac{1}{t}{\cal M}[:\exp(t F^\dagger_{\alpha'}F_{\alpha}):].
\end{equation}

We can apply the RO method to the insertion of two-body operators in a similar manner.  We have the formula
\begin{equation}
    {\cal M}[F^\dagger_{\alpha'_1}F_{\alpha_1}F^\dagger_{\alpha'_2}F_{\alpha_2}] = \lim_{t_1\rightarrow \infty} \lim_{t_2\rightarrow \infty}\frac{1}{t_1t_2}{\cal M}[:\exp(t_1 F^\dagger_{\alpha'_1}F_{\alpha_1}+t_2 F^\dagger_{\alpha'_2}F_{\alpha_2}):].
\end{equation}
We see that the number of amplitudes does not grow with the number of one-body operators, $k$.
}

The RO method can be used entirely by itself or it can be used in some combination of Jacobi formulas and/or numerical derivatives. In this work, we combine the RO method and numerical derivative method for the two-body observables and use the Jacobi formula to handle the perturbative corrections from the higher-order chiral interactions.  As an example, let us consider the normal-ordered two-body density correlation function, $\langle : \hat{\rho}(\bm{n} + \bm{n}') \hat{\rho}(\bm{n}') :\rangle $.
We treat $\hat{\rho}(\bm{n}')$ using the RO formalism \YZ{(in spin and isospin space)} and handle $\hat{\rho}(\bm{n} + \bm{n}')$ using numerical derivatives,
\begin{equation}\label{eq:X18}
        \begin{aligned}
                \mathcal{M}\left[: \hat{\rho}(\bm{n} + \bm{n}') \hat{\rho}(\bm{n}') :\right] 
                =  \lim_{\substack{\epsilon \rightarrow 0 \\ t \rightarrow \infty}} \sum_{i j} 
                \frac{\mathcal{M}\{:\exp[t\hat{\rho}_{ij}(\bm{n}')+\epsilon \hat{\rho}(\bm{n}+\bm{n}')]:\} 
                -
                \mathcal{M}\{:\exp[t\hat{\rho}_{ij}(\bm{n}')]:\} }{t\epsilon} .
        \end{aligned}
\end{equation}
The treatment of the perturbative corrections to 
$\mathcal{M}\{:\exp[t\hat{\rho}_{ij}(\bm{n}')+\epsilon \hat{\rho}(\bm{n}+\bm{n}')]:\}$ and $\mathcal{M}\{:\exp[t\hat{\rho}_{ij}(\bm{n}')]:\}$ from the high-fidelity interactions are handled using Jacobi formulas.

\subsection{RPA calculations of neutrino inverse mean free paths calibrated by lattice results}

\ZL{
Comparing to the virial expansion which has difficulties in achieving convergence at high densities, the NLEFT calculations calculate the structure factors with uncertainty quantification at all densities and serve as an ideal benchmark for calibrating the RPA calculation of neutrino opacities in both CCSNe and binary neutron star mergers. In the following, we perform the first Bayesian inference of Skyrme models using {\it ab initio} lattice calculations as constraints. 

In our Bayesian inference, the posterior distribution of quantity $Q$ is:
\begin{equation}\label{eq:postorig}
    P^{post}_Q(q) = \int \delta[Q(\{p\})-q] {\cal{L}}(\{p\}) P_{\mathrm{prior}}(\{p\})~d\{p\},
\end{equation}
where $q$ is a specific realization of $Q$ and $\{p\}$ are parameters in Skyrme models. The likelihood $\cal{L}$ is written as 
\begin{equation}
    {\cal{L}} = \prod_k \exp \left\{ - \frac{\left[m_k(\{p\})-d_k\right]^2}{2 \sigma^2_k} \right\},
\end{equation}
where $m_k(\{p\})$ denotes the $k$th constraint in the Bayesian inference of Skyrme models, $d_k$ is the mean and finally $\sigma_k$ is the uncertainty of this constraint. In this case, the constraints include the charge radius and the binding energy of three representative nuclei ($^{48}\mathrm{Ca}$, $^{90}\mathrm{Zr}$, $^{208}\mathrm{Pb}$), the charge and the weak form factors of $^{48}\mathrm{Ca}$ and $^{208}\mathrm{Pb}$ measured by CREX and PREX, and finally, the $S_v$ and $S_a$ calculated by NLEFT at 7 different densities. Given the posterior distribution of Skyrme models, we are able to calculate the probability distribution of structure factors based on the RPA methods explained in section H. 

Since the axial current interaction contributes approximately $75\%$ of the neutrino opacity and has relatively large uncertainties, in the following we focus on the structure factors and the neutrino inverse mean free path in the axial current channel. We present the posterior distribution of $S_a$ and $\mathrm{IMPF}_{ax}$ with (the grey band) and without (the orange band) NLEFT constraints, and the $\mathrm{IMPF}_{ax}=\frac{3}{4}G^2_\mathrm{F}g^2_aE^2_\nu nS_a(n)/\pi$. As one may easily observe in Fig.~\ref{fig:X5}, comparing to the probability distribution without NLEFT constraints, the structure factor $S_a$ and the axial neutrino inverse mean free path constrained by lattice calculations significantly decreases and their uncertainties are much smaller.  \begin{figure}[h] 
        \centering 
        \includegraphics[width=0.45\textwidth]{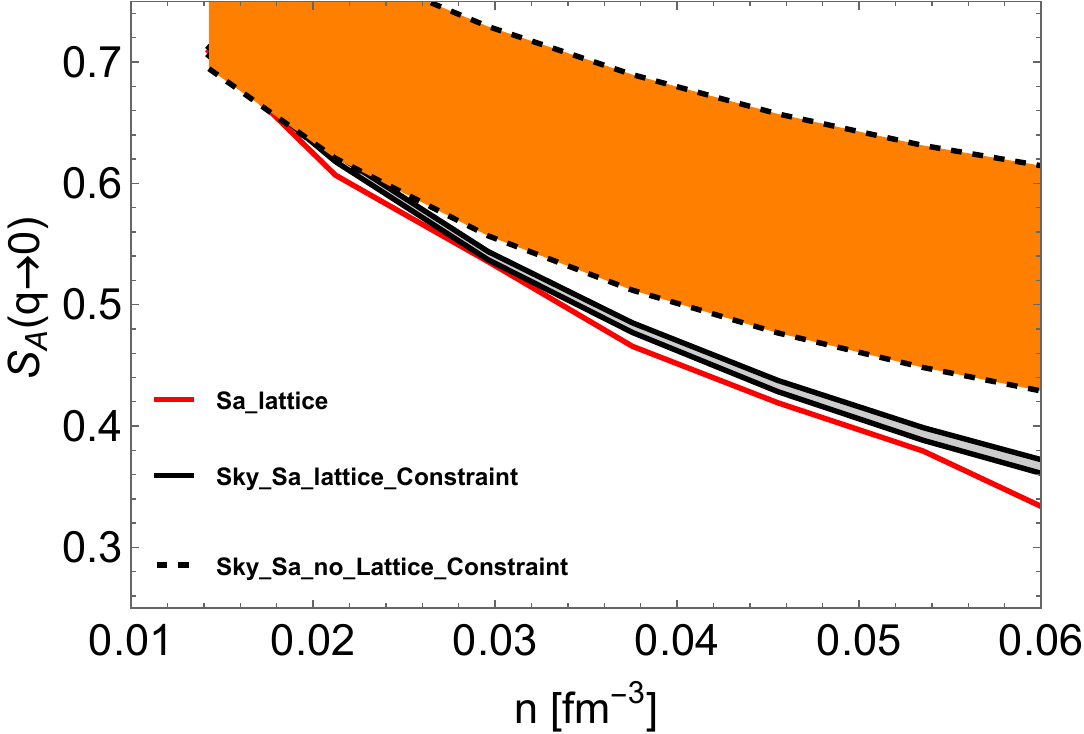}
        \includegraphics[width=0.45\textwidth]{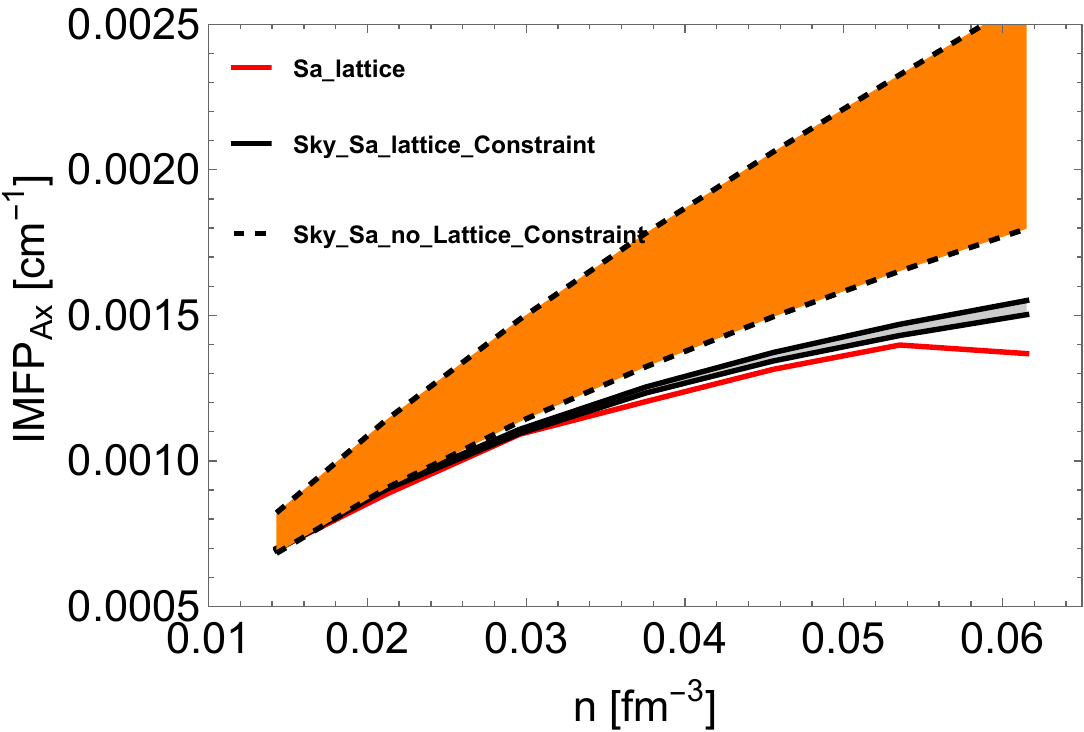} 
        \caption{\label{fig:X5} The axial current structure factors as well as the neutrino inverse mean free path contributed by the axial current reactions. The orange and the grey band represent results coming from Bayesian inference without and with NLEFT constraints, and the red solid curve represents the NLEFT calculations. }
\end{figure} 


\begin{table}[h]
\caption{\label{tab:X2} The mean and uncertainties of the charge radii and binding energies per nucleon of $^{48}\mathrm{Ca}$, $^{90}\mathrm{Zr}$, $^{208}\mathrm{Pb}$ from Bayesian inference. The values in parentheses are experimental data. }
\centering{}%
\setlength{\tabcolsep}{0.2cm}
\begin{tabular}{ccc}
\hline
\hline
   & $R_{ch}$ [fm]  & Binding energy per nucleon [MeV] \tabularnewline
\hline
 $^{48}\mathrm{Ca}$ & $3.43\pm0.012 ~~(3.48)$ 	& $9.93\pm0.23 ~~(8.66)$ \tabularnewline
 $^{90}\mathrm{Zr} $& $4.20\pm0.013 ~~(4.27)$ & $9.69\pm 0.18 ~~(8.71)$  \tabularnewline
 $^{208}\mathrm{Pb}$ & $5.50\pm0.015 ~~(5.5)$ & $8.20\pm0.14 ~~~(7.87)$ \tabularnewline
\hline
\hline
\end{tabular}
\end{table}

From Tab.~\ref{tab:X2}, we find that after including the NLEFT constraints, the Skyrme models can still reasonably describe the charge radii of various nuclei. However, the binding energies predicted by these models are slightly higher than the experimental measurements. This may indicate a minor tension between the description of nuclei binding energies and the axial current structure factors, which will be studied in our future work.

Finally, we have studied the Pearson correlations between the axial structure factors of pure neutron matter and the axial structure factors of dense matter at finite proton fractions, which are calculated based on the Skyrme models and the RPA method. As one may observe in Fig.~\ref{fig:X6}, the correlations between the $S_a$ of pure neutron matter and the $S_a$ of matter at low $Y_e$ are strong at all densities, which indicate that NLEFT lattice calculations performed at pure neutron matter may strongly constrain the calculations of structure factors of neutron-rich matter that are relevant to the environment in CCSNe and binary neutron star mergers, in the framework of NLEFT-calibrated RPA method.  In the future, lattice calculations with nonzero proton fraction will also be performed to provide additional calibration tools of RPA calculations and CCSNe simulations.

\begin{figure}[h] 
        \centering 
        \includegraphics[width=0.45\textwidth]{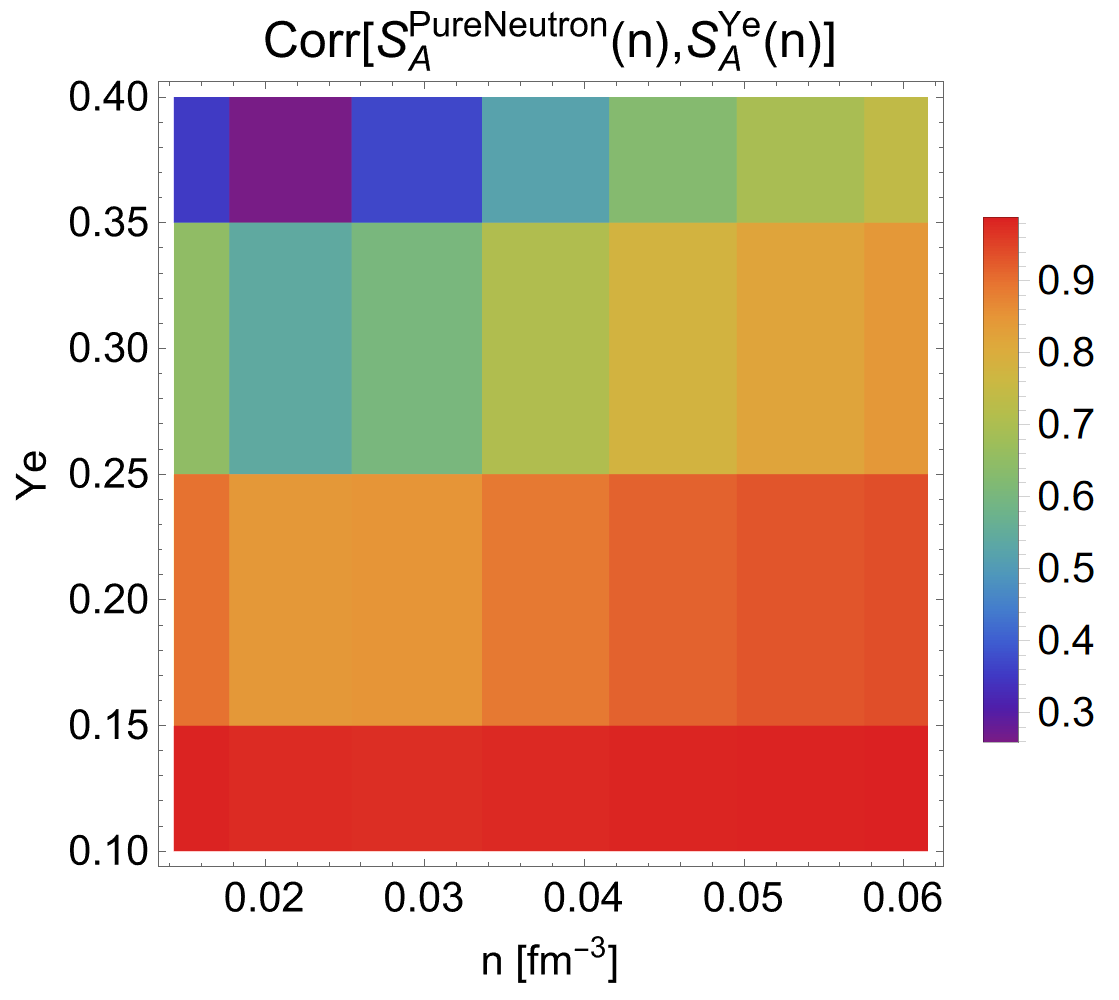}
        \caption{\label{fig:X6} Pearson correlation between the axial current structure factors of pure neutron matter and the axial current structure factors of dense matter at finite proton fractions }
\end{figure}   

}

\subsection{Grand canonical ensemble benchmarks for a free Fermi gas}
\paragraph{}As noted in the main text, observables in the grand canonical ensemble (GCE) can be calculated using a series of canonical ensembles (CE) with a weight distribution $w_{N}=e^{\beta \mu_G N} \frac{Z(\beta,T)}{\mathcal{Z}(\beta,\mu_G)}$.
In the thermodynamic limit, $w_N$ will have a Gaussian dependence near its maximum.  We take the derivative of $w_N$ with respect to $N$ and obtain,
\begin{equation}
        \begin{aligned}
        \frac{\partial}{\partial N} \omega_N & =\omega_N \frac{\partial}{\partial N}\left[\mu_G N \beta-\beta F(N)\right] \\
        & =\omega_N \beta\left[\mu_G-\mu(N)\right].
        \end{aligned}
\end{equation}
This means that the Gaussian function will have the maximum when $\mu_G = \mu(N)$, where $\mu(N)$ is the $N$-particle CE chemical potential.
In our lattice calculation, the CE chemical potential is calculated using the Widom insertion method \cite{Widom1963, Binder1997, Lu2020}.

\begin{figure}[!ht] 
        \centering 
        \includegraphics[width=0.55\textwidth]{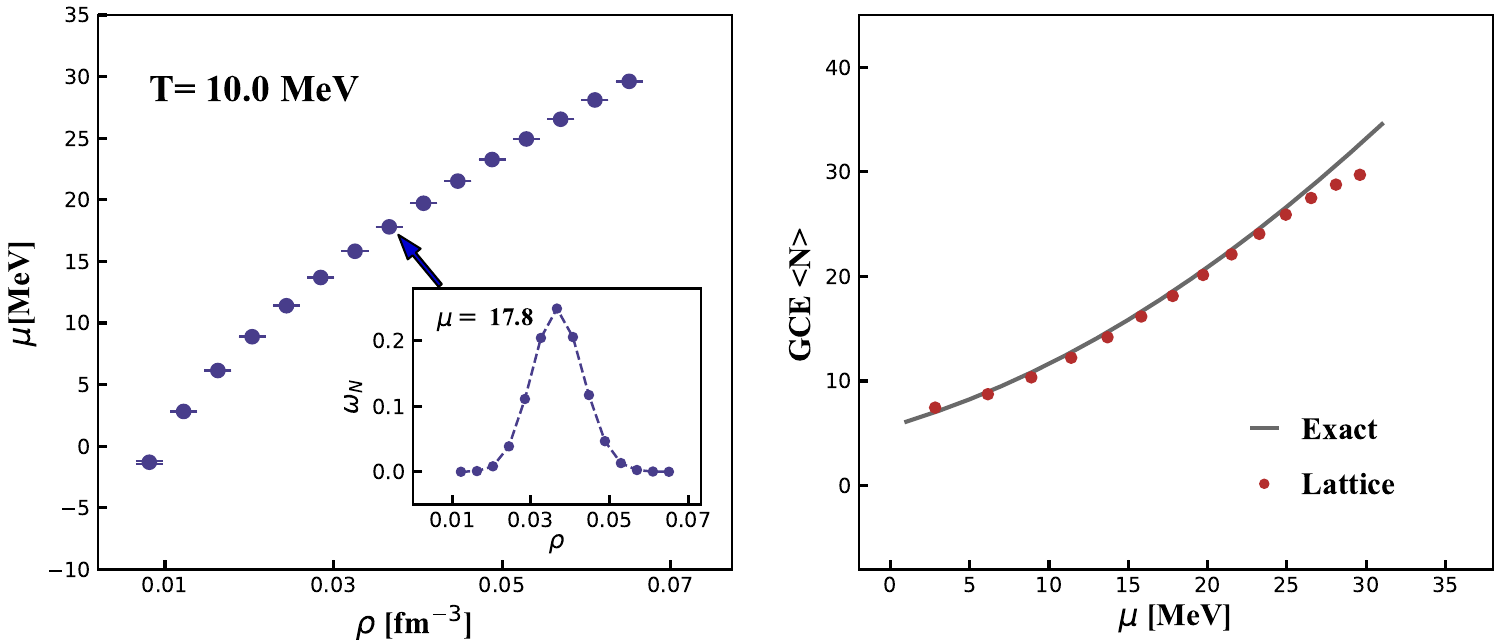} 
        \caption{\label{fig:S1} Free Fermi gas chemical potentials are shown in the left panel.
        The insert figure shows the weight distribution $w_N$ of CE systems for a GCE system at $\mu=17.8$ MeV. 
        The right panel shows the comparison of the calculated GCE particle number (red dots) and the analytical calculation (solid line).} 
\end{figure}

We can benchmark these calculations for a free Fermi gas of neutrons.  In the grand canonical ensemble, the particle number of free Fermi gas 
can be obtained from the integral of the level density with the Fermi-Dirac distribution,
\begin{equation}\label{eq:FermiGas}
    \begin{aligned}
        \int_0^\Lambda \frac{\rho(\epsilon)}{1+e^{\beta(\epsilon - \mu)}} d \epsilon = N,
    \end{aligned}
\end{equation}
where $\Lambda = (\pi/a)^2/(2m)$ is the energy cutoff imposed by the lattice spacing $a$ \cite{Lu2020}.
Its level density can be obtained using
\begin{equation}
        \begin{aligned}
           \rho(E)=\frac{d N}{d E}
           &=\frac{1}{\pi^2\hbar^3} m V \sqrt{2mE}.
        \end{aligned}
\end{equation}
In Fig. \ref{fig:S1}, we show the free neutron gas results for an $L^3=6^3$ lattice box at $T=10$ MeV. 
The left panel shows the CE chemical potential at different densities and the inset shows the CE weight distribution for the GCE with $\mu_G=17.8$ MeV.
In the right panel, we show the expectation values of the GCE particle number and benchmark them against the analytical solution in Eq.~\eqref{eq:FermiGas}.
We find a nice agreement between calculated GCE particle numbers and that from the analytical solutions. 
The small deviation in the last few points can be resolved by including more CE systems. 



\subsection{Perturbation theory corrections to the static structure factors}
To verify the corrections from the first-order perturbation theory to the structure factors, we perform lattice calculations with a simple Hamiltonian of the form
\begin{equation}
\begin{aligned}
        H = K + \frac{1}{2} c_{\rm{SU4}} \sum_{\bm{n}}: \tilde{\rho}^2(\bm{n}) :,
\end{aligned}    
\end{equation}
where $K$ is the kinetic term.  This simple Hamiltonian allows us to do fully non-perturbative lattice calculations without any sign oscillations.  For our benchmark calculations, we introduce a parameter $x$ that divides the original Hamiltonian $H$ into a non-perturbative part (see also Ref.~\cite{Lahde:2015ona}),
\begin{equation}
        H_0 = K + (1-x)\times \frac{1}{2} c_{\rm{SU4}} \sum_{\bm{n}}: \tilde{\rho}^2(\bm{n}) :,  
\end{equation}
and a perturbative correction,
\begin{equation}
        H_1 = x\times \frac{1}{2} c_{\rm{SU4}} \sum_{\bm{n}}: \tilde{\rho}^2(\bm{n}) :.
\end{equation}
We note that for any value of $x$, $H = H_0 + H_1$, and the parameter $x$ controls the size of the $H_1$ term.

\begin{figure}[h] 
        \centering 
        \includegraphics[width=0.5\textwidth]{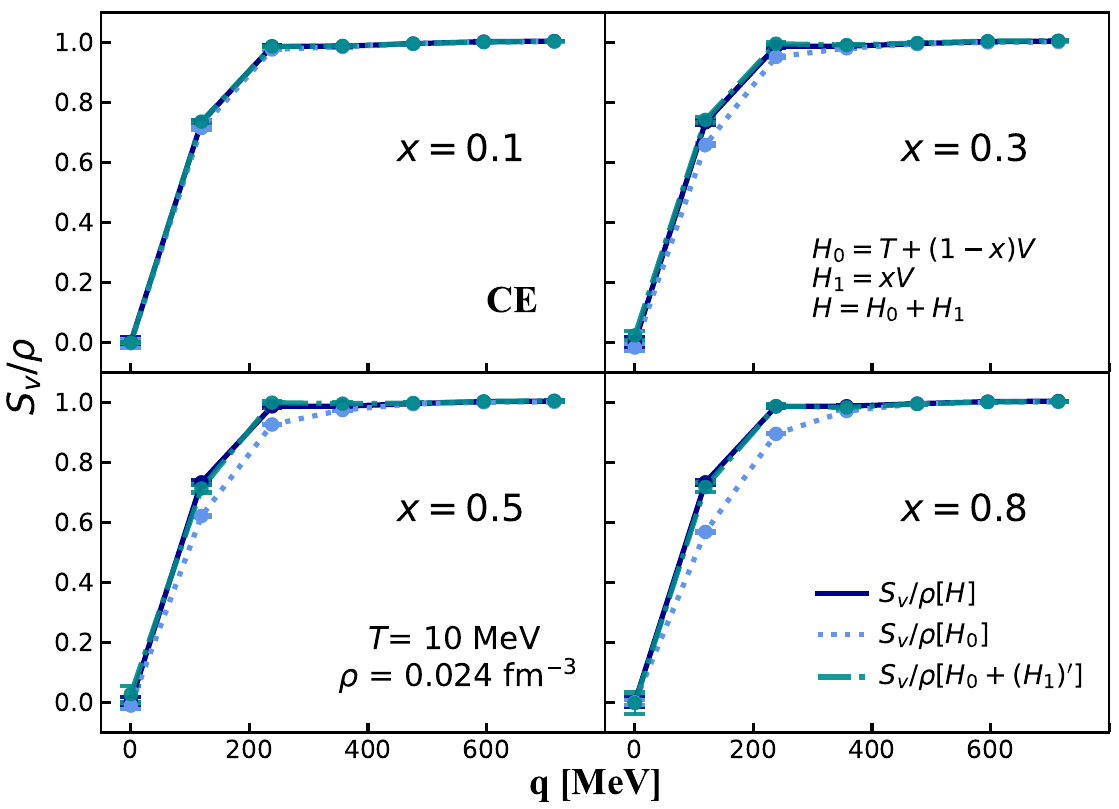} 
        \caption{\label{fig:X2} Calculated CE static vector structure factor $S_{\rm{v}}$ with the full Hamiltonian $H$ (solid line), 
        with $H_0$ only (dotted line), and with $H_0 + H_1$ where $H_1$ is included at first-order in perturbation theory (dot-dashed line).}
\end{figure}

\begin{figure}[h] 
        \centering 
        \includegraphics[width=0.5\textwidth]{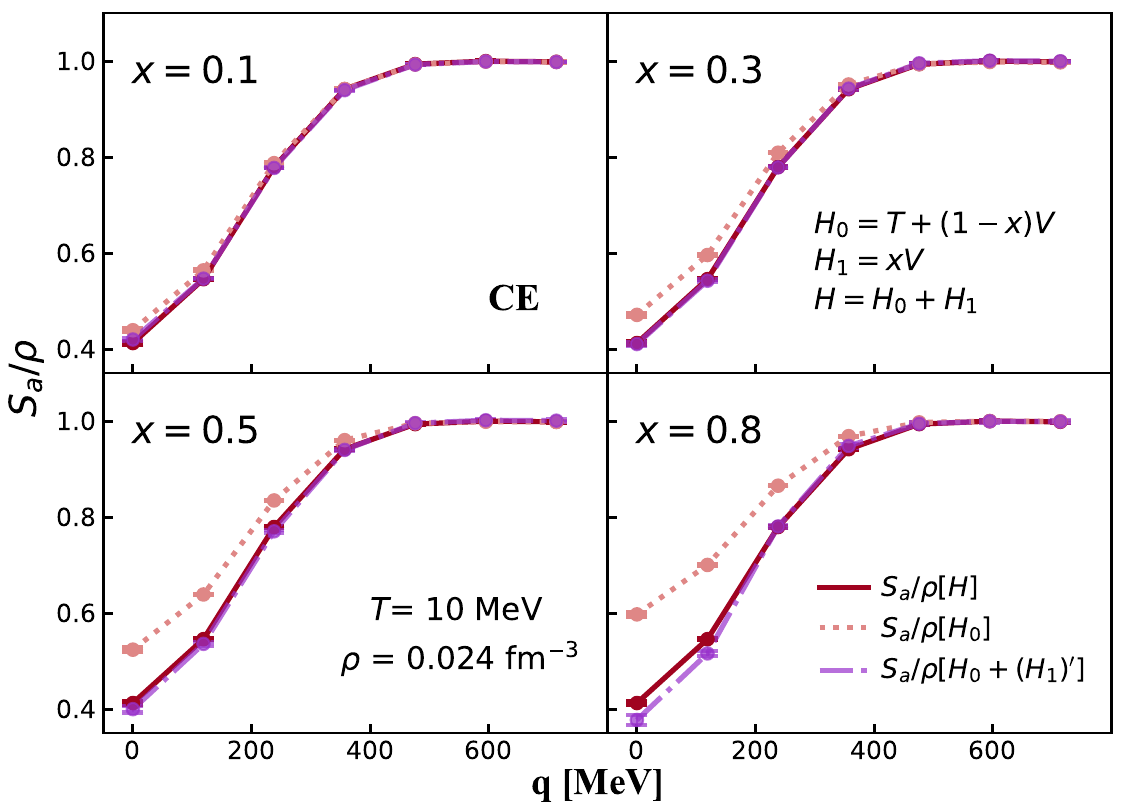} 
        \caption{\label{fig:X3} Calculated CE static axial structure factor $S_{\rm{a}}$, with the full Hamiltonian $H$ (solid line), 
        with $H_0$ only (dotted line), and with $H_0 + H_1$ where $H_1$ is included at first-order in perturbation theory (dot-dashed line).}
\end{figure}
In Fig.~\ref{fig:X2} and Fig.~\ref{fig:X3}, we show the momentum-dependence of $S_{\rm v}$ and $S_{\rm a}$ at $T=10$ MeV in the canonical ensemble.
The first-order perturbation theory calculation for $S_{\rm{v}}$ and $S_{\rm{a}}$ are benchmarked against results obtained with non-perturbative calculations using the full Hamiltonian $H = H_0 + H_1$.
We see that $S_{\rm{v}}$ and $S_{\rm{a}}$ for $H_0$ (dotted lines) are diverging from the ``full'' Hamiltonian results (solid lines) as we increase $x$.  Nevertheless, the perturbation theory results $S_{\rm{v}/\rm{a}}[H_0+(H_1)']$ (dot-dashed lines) agree quite nicely with the ``full'' Hamiltonian results.  For the axial structure factor, only a small discrepancy happens at $x=0.8$ where $H_0$ is quite different from $H$.

\subsection{Perturbation theory corrections to the chemical potential}
We use the Widom insertion method \cite{Lu2020} to compute differences in the free energy and calculate the chemical potential.  In Table~\ref{tab:X1} we show results at different orders in the chiral expansion at $T=10$ MeV in the canonical ensemble for a range of densities.  $H^S$ corresponds to the Hamiltonian of Eq.~\ref{eq:HS}, which is used for non-perturbative calculations. $Q^0$, $Q^2$ and $Q^4$ represents different orders in the $\chi$EFT expansion.  WFM(N3LO) includes wave function matching corrections at N3LO.  
It should be mentioned that $H^S$ is similar to $Q^0$ which contains one-pion exchange and two S-wave contact interactions.
However, to reduce the sign problem, $H^S$ has a softer regulator for the one-pion exchange and does not include the counterterm of Eq.~\eqref{eq:OPEcounter}. See Ref.~\cite{Elhatisari2022} for more details.
The differences between $H^S$ and high-fidelity $\chi$EFT interactions are treated in perturbation theory.

We observe that the chemical potentials at different orders in $\chi$EFT along with non-perturbative $H^S$ all agree with each other within stochastic error bars.  This is most likely due to the fact that all of the interactions have approximately the same S-wave phase shifts.  Given the negligible variation in the chemical potential from the perturbative corrections, we have simplified our lattice calculations by neglecting these small differences in the chemical potentials and using the $H^S$ results.

\begin{table}[h]
\caption{\label{tab:X1} Perturbation calculation of chemical potentials $\mu$ at $T=10$ MeV.  $H^S$ corresponds to the Hamiltonian of Eq.~\ref{eq:HS}, which is used for non-perturbative calculations. $Q^0$, $Q^2$ and $Q^4$ represents different order in the $\chi$EFT expansion.  WFM(N3LO) includes the corrections from wave function matching at N3LO, corresponding to Eq.~\ref{eq:wfm}.}
\centering{}%
\setlength{\tabcolsep}{0.2cm}
\begin{tabular}{cccccc}
\hline
\hline
 density [fm$^{-3}$]  & $H^S$  & $Q^0$  & $Q^2$  & $Q^4$  & WFM(N3LO)  \tabularnewline
\hline
 $0.012$ & $-4.732(24)$	& $-4.706(24)$ & $-4.709(24)$ & $-4.715(24)$ & $-4.717(24)$ \tabularnewline
 $0.016 $& $-2.626(25)$ & $-2.593(25)$ & $-2.595(25)$ & $-2.605(25)$ & $-2.609(25)$ \tabularnewline
 $0.020$ & $-0.953(26)$ & $-0.911(26)$ & $-0.916(26)$ & $-0.929(26)$ &  $-0.934(26)$ \tabularnewline
 $0.024$ & $0.440(31)$ & $0.485(31)$ & $0.481(31)$ & $0.460(31)$ & $0.456(31)$ \tabularnewline
 $0.028$ & $1.708(39)$ & $1.767(38)$ & $1.762(38)$ & $1.736(38)$ & $1.744(39)$ \tabularnewline
 $0.033$ & $2.933(54)$ & $2.992(54)$ & $2.987(55)$ & $2.953(56)$ & $2.990(59)$
 \tabularnewline
\hline
\hline
\end{tabular}
\end{table}
        
\subsection{RPA and Virial structure factors}
In this subsection, we briefly summarize the needed formulas for calculating RPA $S_{\rm{v}}(q)$ and $S_{\rm{a}}(q)$. The dynamic vector (axial) structure factor of pure neutron matter based on RPA calculations is written as
\begin{equation}
S_{\rm RPA}(q_0,q)=\dfrac{2~\mathrm{Im}\Pi_{\rm RPA}}{1-\exp[-q_0/T]},\nonumber 
\end{equation}
where $q_0$ is the transferred energy and $q$ is the transferred momentum by scattered neutrinos. The $\Pi_{RPA}$ is the polarization function of pure neutron matter, and is calculated given mean-field neutron matter polarization functions $\Pi_0$,
\begin{equation}
\label{eq:NCrpaPIvec}
\Pi_{\rm RPA}=\dfrac{\Pi_0}{1-V_{\rm res} \Pi_0} \, ,
\end{equation}
where $V_{\rm res}$ is the (spin-dependent) residual interactions between neutrons. The detailed expression of $\Pi_0$ is provided in Ref.~\cite{Lin:2022lug}. In vector current channel, $V_{\rm res}=f_{nn}$. In axial current channel, $V_{res}=g_{nn}$. The $f_{nn}$ and $g_{nn}$ are Landau-Migdal parameters and their detailed expression in terms of Skyrme parameters was carefully derived in Ref.~\cite{Lin:2022lug}. Given the dynamic structure factor $S_{\rm RPA} (q_0, q)$, we obtain the unitless static structure factor $S_{\rm RPA}(q)$ from
\begin{equation}
    S_{\rm RPA}(q)=\frac{1}{2\pi n}\int S_{\rm RPA} (q_0, q) dq_0,
\end{equation}
where $n$ is the number density of pure neutron matter.

For the virial expansion up to the 4th order, the number density of pure neutron matter is written as 
\begin{equation}\label{eq:virialden}
    n=\frac{2}{\lambda^3}(z+2z^2b_2+3z^3b_3+4z^4b_4).
\end{equation}
The virial coefficients $b_n$ for the unitary Fermi gas have been determined both theoretically and experimentally from ultracold gas experiments. We can use \ref{eq:virialden} to obtain the fugacity $z$ as a function of density. Given the fugacity, the unitary-limit virial static structure factors in long wavelength limit can be written as 
 \begin{equation}\label{eq:virialsv4}
     S_{\rm v}(q\rightarrow0)=\frac{1+4zb_2+9z^2b_3+16z^3b_4}{1+2zb_2+3z^2b_3+4z^3b_4} ~,
 \end{equation}
and 
\begin{equation}\label{eq:virialsa4}
     S_{\rm a}(q\rightarrow0)=\frac{1+4zb^0_2+z^2(8b^0_3+b_3)+z^3(16b^0_4+4b_{3,1})}{1+2zb_2+3z^2b_3+4z^3b_4} ~.
 \end{equation}
 The virial coefficients used in Eq.~\eqref{eq:virialsv4} and \eqref{eq:virialsa4} are the same as in~\cite{Lin2017}.  $b^0_n$ denotes the virial coefficents for the free Fermi gas, and $b_{3,1}$ is the 4th order coefficient in the unitary limit for three neutrons of the same spin and one neutron of the opposite spin.

 If we apply the virial expansion to realistic neutron matter up to 2nd order, we have
 \begin{equation}\label{eq:virialsv2}
     S_{\rm v}(q\rightarrow0)=1+\frac{1+4z^2b_2}{\lambda^3 n} ~,
 \end{equation}
 and 
 \begin{equation}\label{eq:virialsa2}
     S_{\rm a}(q\rightarrow0)=1+\frac{4}{\lambda^3}\frac{z^2 b_a}{n} ~,
 \end{equation}
 where $\lambda=\sqrt{2\pi/m T}$ is the thermal wavelength and the virial coefficients are extracted from nucleon-nucleon scattering phase shifts. The virial coefficients here are the same as those in \cite{Horowitz2017}. Note that by setting the 3rd and the 4th order virial coefficients to zero, Eq.~\eqref{eq:virialsv4} and \eqref{eq:virialsa4} reduce to Eq.~\eqref{eq:virialsv2} and Eq.~\eqref{eq:virialsa2}.

\begin{figure}[h] 
        \centering 
        \includegraphics[width=0.45\textwidth]{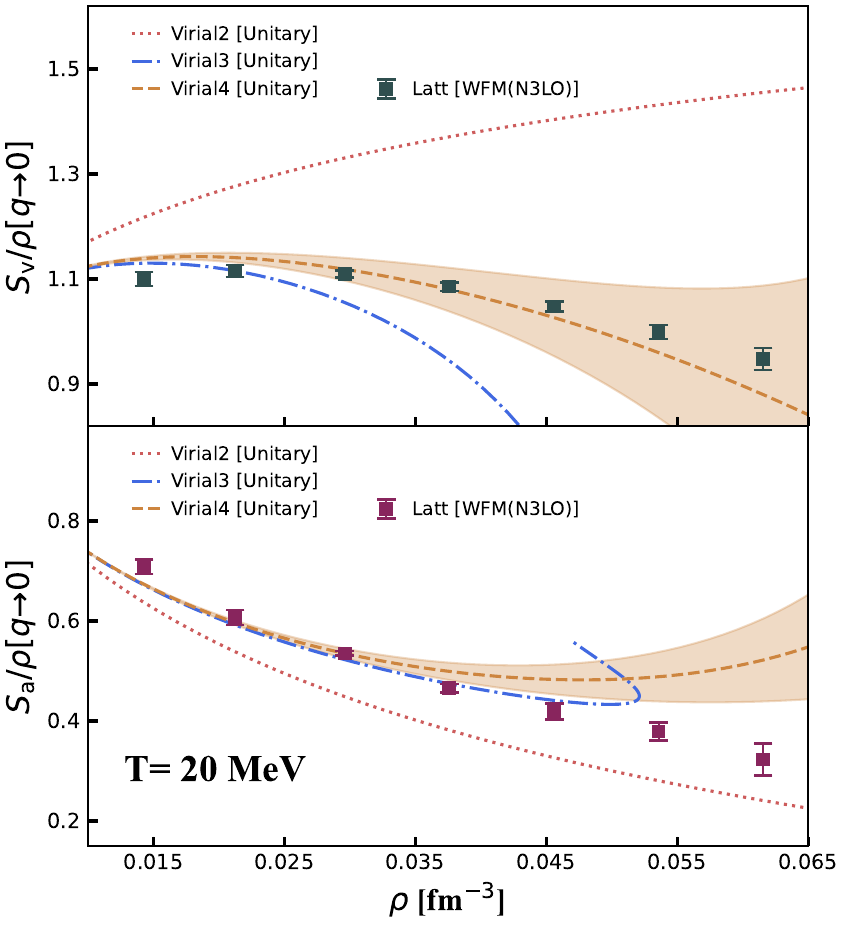} 
        \caption{\label{fig:X4} Comparison of structure factors between unitary virial and NLEFT calculations. We show results at the 2nd, 3rd, and 4th orders in the virial expansion for the unitary limit.  The error band reflects the theoretical uncertainty in the fourth-order virial coefficients. WFM(N3LO) stands for lattice results with $\chi$EFT interactions at N3LO.}
\end{figure}



\paragraph{} \YZ{In the main text, we found that the Virial4 [Unitary] results intersect with the lattice results for the vector and axial static structure factors near density $0.030$~fm$^{-3}$. 
In Fig.~\ref{fig:X4}, we plot unitary results for the virial expansion at 2nd, 3rd and 4th orders. While the order-by-order convergence at low densities is good, the convergence for $\rho $ larger than $0.03$~fm$^{-3}$ is much slower, especially for the axial structure factor. In additional to the virial truncation error, there are also sizable theoretical uncertainties arising from the fourth-order virial coefficients, $b_4=0.047(18)$ and $b_{3,1}=0.170(13)$~\cite{PhysRevLett.105.070402, Lin2017}.  These contribute an additional large uncertainties at higher densities.}

\paragraph{}
\paragraph{}
\paragraph{}
\paragraph{}
\paragraph{}
\paragraph{}
\paragraph{}

\bibliography{References}
\bibliographystyle{apsrev}